\documentclass[twocolumn]{article}

\usepackage{ amsmath, amssymb, multicol, graphicx}%pupr_paper,
\usepackage{flushend}%\usepackage{Times}

\begin{document}

\title{Simulation of a Hyperbolic Field Energy Analyzer}
\author{Angel Gonzalez-Lizardo, Ernesto Ulloa}

\date{}
\maketitle

\begin{abstract}
Energy analyzers are important plasma diagnostic tools with applications in a broad range of disciplines including molecular spectroscopy, electron microscopy, basic plasma physics, plasma etching, plasma processing, and ion sputtering technology. The Hyperbolic Field Energy Analyzer (HFEA) is a novel device able to determine ion and electron energy spectra and temperatures. The HFEA is well suited for ion temperature and density diagnostics at those situations where ions are scarce. A simulation of the capacities of the HFEA to discriminate particles of a particular energy level, as well as to determine temperature and density is performed in this work.
The electric field due the combination of the conical elements, collimator lens, and Faraday cup applied voltage was computed in a well suited three-dimensional grid. The field is later used to compute the trajectory of a set of particles with a predetermined energy distribution. The results include the observation of the particle trajectories inside the sensor, the comparison of the input energy distribution to the energy distribution of the particles captured by the Faraday cup, and the IV characteristic at the Faraday cup, using the voltage sweep at the conical elements as the abscissa.
\end{abstract}

\section{Introduction}
Energy analyzers are important plasma diagnostic tools with applications in a broad range of disciplines including molecular spectroscopy, electron microscopy, basic plasma physics, plasma etching, plasma processing, and ion sputtering technology.
Retarding Potential Analyzers (RPAs) are diagnostic tools generally using a series of electrostatic grids to selectively repel some particles in order to detect the ion energy distribution. Typical construction of an RPA comprises three electrostatically-biased mesh grids and an optional fourth grid. Grids are aligned inside a housing with a detector placed behind them. In this work, a simulation of a  grid-less energy analyzer is presented. Devices similar  to the one presented in this work have been used for detection of low energy particles \cite{SSH1976}, focusing of space-charge-dominated electron beams \cite{CZVRWHKBO2004}, ionospheric induced perturbations caused by the space shuttle \cite{JGRA:JGRA8211}, and other applications. Additionally, single probes which can measure multiple parameters simultaneously and can operate at relatively high temperatures have uses in the analysis of high temperature plasmas generated in fusion research devices. The Hyperbolic Field Energy Analyzer (HFEA) is a device able to determine ion and electron energy spectra and temperatures. \cite{Le1989} At low plasma densities, Langmuir probes are well suited for measuring electron temperatures, but ion temperatures may be a challenge \cite{opac-b1133704}. The HFEA is well suited for ion temperature and density diagnostics at those situations where ions are scarce. The HFEA consists of three main components: a plane parallel disc with interchangeable diameter orifices for rejection of electrons and collimation of the ion beam, a dual conical lens for particle energy selection and beam focusing, and a Faraday Cup as the particle collector.

\begin{figure}[htb!]
   %Requires \usepackage{graphicx}
 \includegraphics[width=\columnwidth]{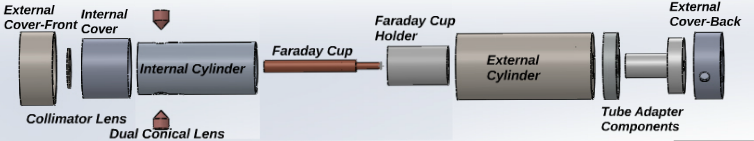}\\
 \caption{ HFEA Components }\label{HFEA1}
  \end{figure}
An exploded view of the HFEA parts is shown in Figure \ref{HFEA1}. According to Leal-Quir\'os \cite{Le1989}, the conical lens constitute a hyperbolic cross-section lens due to the equipotential surfaces in the vicinity of a diaphragm, which are hyperboloids of revolution. A saddle point crossed by straight field lines forms in the center of the device. An angle of $54^\circ-44$'  between the straight lines and the axis at the saddle point was used for designing the conical lenses. If a potential $V_R$ is imposed on the saddle point, only paraxial ions with energy greater than $q V_R$ can pass through the lens into the Faraday Cup.

It is known that small systems of a few thousand particles can simulate accurately the collective behavior of real plasmas \cite{opac-b1121477}. With this in mind, a simulation of the capacities of the HFEA to discriminate particles of a particular energy, as well as to determine temperature and density is performed in this work. The electric field created by the combination of the conical elements, collimator lens, and Faraday cup applied voltage was computed in a three-dimensional grid and used to compute the trajectory of a set of particles with a predetermined energy distribution. The results include particle trajectories inside the sensor, the comparison of the input energy distribution to the energy distribution of the particles captured by the Faraday cup, and the Current-Voltage (IV) characteristic at the Faraday cup, using the voltage sweep at the conical elements as the abscissa.
\begin{figure}[htb!]
  \centering
  \includegraphics[width=\columnwidth]{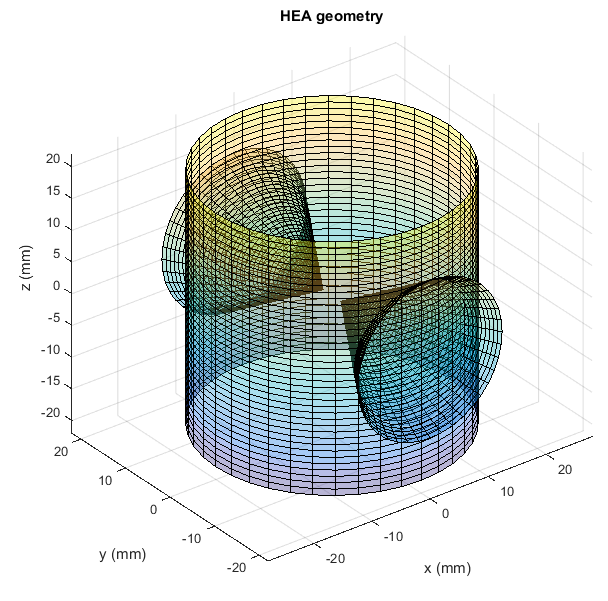}\\
  \caption{HFEA geometry}\label{geometry}
\end{figure}

\section{Methodology}

A matlab simulation of the electric field near the conical lens is performed and analyzed. The geometry of the device was created in matlab (Figure \ref{geometry}) using surfaces to represent the boundary between the different elements of the analyzer.  Figure \ref{geometry} shows the two-cone electrode and the outer cylinder containing the assembly.  Initial voltage conditions were given to each of the elements in order to find the solution of equation \ref{laplace}, with $\Phi$ the electric potential matrix at the 3D space in the vicinity of the conical lens.

\begin{equation}\label{laplace}
  \nabla^2 \Phi = 0,
\end{equation}

This equation was numerically solved for a rectangular grid of $n \times n \times n$  with Dirichlet boundary condition. The potential at the surfaces of the device under study were fixed. The 3D electric field is found by and computing the gradient of the potential at each point of the grid, by

\begin{equation}\label{E}
  \overrightarrow{E} =-\nabla \Phi
\end{equation}

The solution for the electric field was used later on to compute the forces exerted on particles entering the device in a kinematic simulation of their trajectories. This trajectory simulation was performed by assuming the entering particles had a Maxwellian energy distribution. The distribution of velocities of the particles reaching the Faraday cup was obtained and compared with the initial distribution.

Figure \ref{E} shows a graphical representation of the equipotential surfaces in the plane $yz$ of the device. The colored graph shows the lower potential in blueish colors, while the highest potential are closer to the red. The faraday cup voltage can be observed as a yellow oval in top of the $z$ axis.

\begin{figure}[htb!]
  \centering
  \includegraphics[width=\columnwidth]{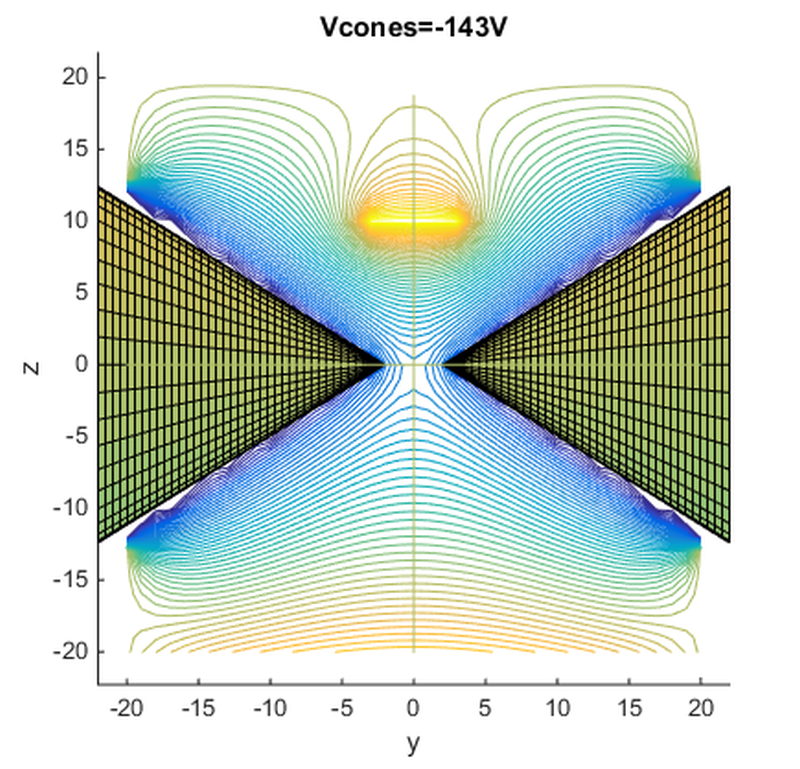}\\
  \caption{Equipotential surfaces for V$_{\textrm{C}}$=-143 V}\label{E}
\end{figure}

\begin{figure}[htb!]
  \centering
  \includegraphics[width=\columnwidth]{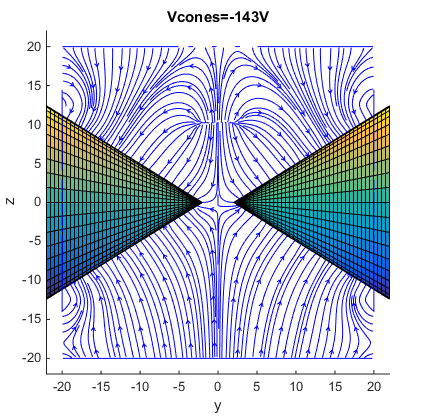}\\
  \caption{Electric Field for V$_{\textrm{C}}$=-143 V}\label{E40}
\end{figure}

Figure \ref{E40} shows the electric field lines obtained for a particular negative voltage applied to the two-cones electrode.

\section{Simulations}
The methodology used in this work is quite simple. The objective was to produce a scaled version of the  IV characteristic obtained in \cite{Le1989}  for electrons and ions. A simulation using the two species in equal number was performed to achieve the goal.

To simulate the effect of applying a ramp voltage from -200 V to 200 V to the two-cones electrode in the device, the 3D electric field produced by all the voltages in the device $\vec{\mathbf{E}}$ was computed for the ramp voltage applied to the  electrode, in steps of one volt. The electric field is computed at discrete points of the space inside the device in a $ n \times n \times n$ grid.  The 3D electric field grid for each level of voltage is stored in a file to be used in the subsequent particle trajectory simulation. Figure \ref{E40} shows a profile of the electric field at the central plane of the device for a particular level of voltage. Figure \ref{E} shows a profile of  the equipotential surfaces at the same plane.

When the electric field data for all 401 levels of voltage was obtained, a simulation of particles being subject to the electric field for each step of voltage was performed. A number of particles $n_p$ were defined as entering the device at a random position within the collimator lens input hole. The velocity of the entering particles was defined as obeying a normal distribution on the $z$ axis and zero velocity in the other two directions (this condition can be easily changed). For the case of simulation of ions and electrons and ions together, only hydrogen ions that lost one electron were included in the simulation for simplicity. The following classical mechanics equations of motion were used in this simulation:

\begin{eqnarray}
  \vec{a}_p &=& \frac{\vec{E}}{m} \label{a_E}\\
  \vec{\Delta s_p} &=&  \vec{v_p} t_s + \vec{a} t_s^2\\
  s_p &=& s_p + \vec{\Delta s_p} \\
  \vec{v}_p &=& \vec{v}_p + \vec{a} t_s
\end{eqnarray}

Where $\vec{a}_p$ is the particle acceleration, $\vec{E}$ is the three-dimensional magnetic field at the closest grid position from the particle, $\vec{\Delta s_p}$ is the incremental change in particle position, $\vec{v}_p$ is the particle velocity, $t_s$ is the time step selected for calculation, and $s_p$ is the current position of the particle.

The electric field used for the acceleration calculation was approximated to the electric field at the nearest point of the grid from the position of the particle $\vec{E}_{\gamma}$. Hence for a particle $i$, with  $i=1 \dots n_p$ with $n_p$ the total number of particles in the simulation, its acceleration at the position $(x_i^k, y_i^k, z_i^k)$, with $k$ representing the $k^{th}$ step given by the particle in its trajectory, is computed as \cite{opac-b1121477}:

\begin{equation}\label{a_ik}
  \vec{a} _i^k= \frac{q_i \vec{E}_{\gamma}}{m_i}
\end{equation}

The displacement of the particle $i$ in step $k$ is computed using

\begin{equation}\label{ds_k}
\Delta s_i^{k} = \vec{v}_i^{(k-1)} t_s + \vec{a}^k_i \frac{t_s^2}{2}
\end{equation}

Thus, the absolute position of the particle $i$ in step $k$ is calculated as

\begin{equation}\label{s_p}
  s_i^{k+1} = s_i^{k}+ \Delta s_i^{k}
\end{equation}

and the velocity vector is computed as

\begin{equation}\label{vp_k}
  \vec{v}_p^k = \vec{v}_p^{k-1}+ \vec{a}^k_i t_s
\end{equation}

The position of each particle in the simulation is compared to the coordinates of all the surfaces in the device to detect the end of its trajectory. When the particle touches any surface, no further calculation of position, velocity and acceleration is performed. The particles that hit the Faraday cup input are counted to compute both the current in the Faraday cup and the energy probability distribution of particles at the Faraday cup.

\section{Results}

\begin{figure}[htb!]
  \centering
  \includegraphics[width=\columnwidth]{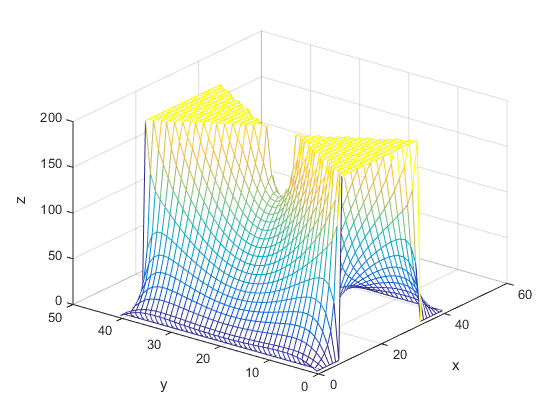}\\
  \caption{Saddle point in the center plane of the voltage}\label{saddle}
\end{figure}

One of the first results being pursued by the simulation was the verification of the formation of a saddle point in the potential generated by the voltage applied to the cone electrodes in the center of the segment connecting the tips of the conical electrode. Figure \ref{saddle} shows the saddle point formed by the voltage magnitude in the center plane for an arbitrary voltage applied to the cones $V_{\textrm{C}}$ and a small voltage applied to the Faraday cup. It can be observed that the voltage barrier is minimum at the center between the two conical electrodes.

\begin{figure}[htb!]
  \centering
  \includegraphics[width=\columnwidth]{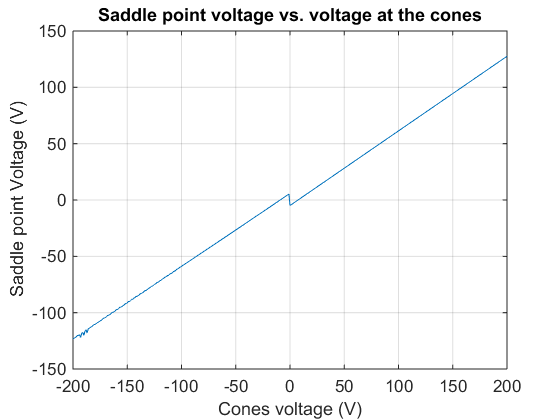}
  \caption{Saddle point voltage vs. $V_{\textrm{C}}$ for Faraday cup Voltage of 67 V, and  collimator voltage of $\pm$10 V, switching voltages polarities to the opposite of the ramp}\label{Vc_vs_saddle}
\end{figure}

Figure \ref{Vc_vs_saddle} shows the saddle point voltage as a function of the $V_{\textrm{C}}$ for the particular of a ramp from -200 V to 200 V, $\pm$67 V at the Faraday cup, and a collimator voltage of $\pm$10 V.  It most be pointed out that the polarity of the Faraday cup and collimator lens voltages  were opposite to the $V_{\textrm{C}}$ sign to obtain this particular plot, explaining the discontinuity at zero.

\begin{figure}[htb!]
  \centering
  \includegraphics[width=\columnwidth]{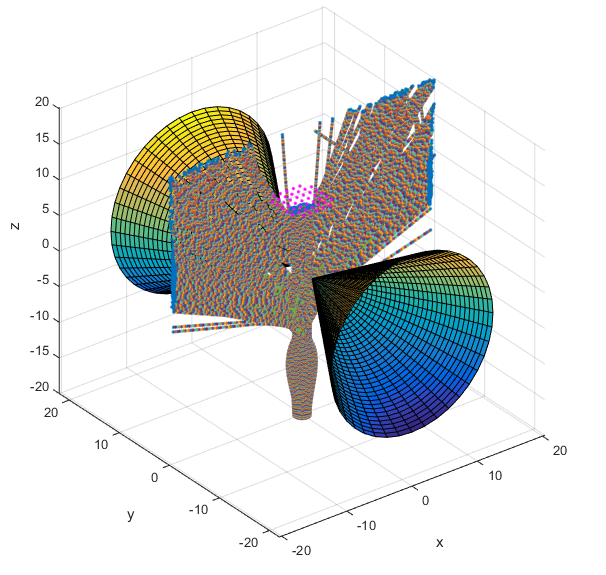}
  \caption{Simulation graphical output for 1000 particles, -200 to 200 V, 67 V and 10 V}\label{simrun}
\end{figure}

\begin{figure}[htb!]
  \centering
  \includegraphics[width=\columnwidth]{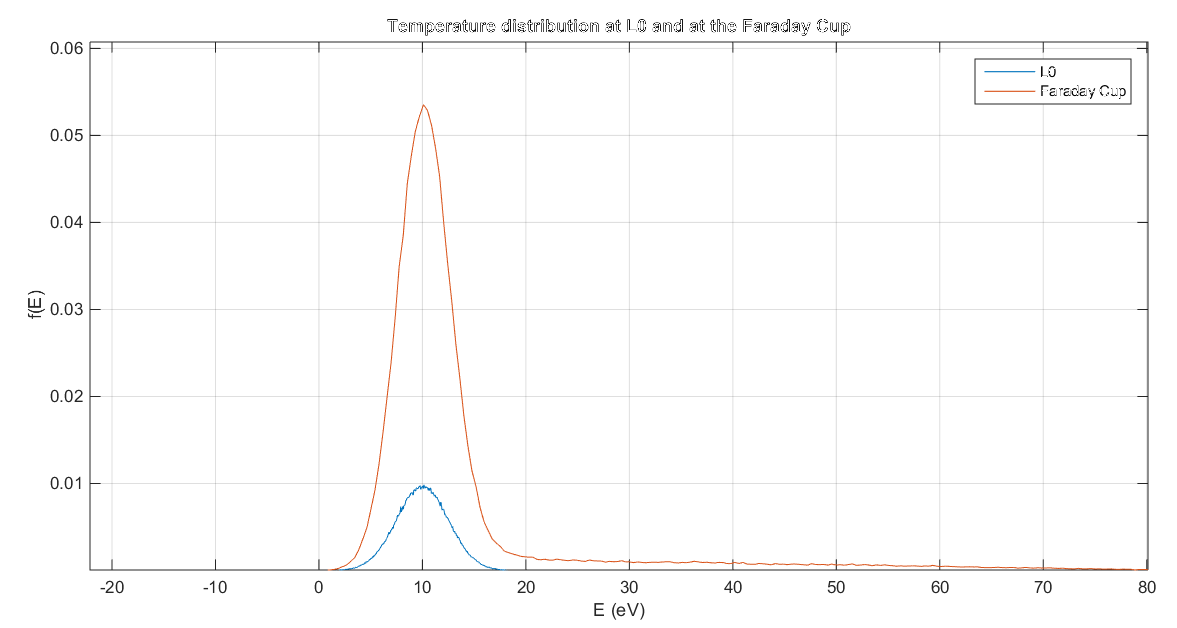}
  \caption{Particle Energy Distribution at Collimator lens and at the Faraday Cup.}\label{Input_and_Output_distributions}
\end{figure}

Figure \ref{simrun} shows the graphic output for a particle simulation using 1,000 particles with a normal energy distribution on the $z$ axis of the simulation, shown in Figure \ref{Input_and_Output_distributions} shows the energy distribution impressed to the particles entering the device. The same number of ions and electrons were included in the simulation to keep the neutrality of the total charge. It may be observed that particles tend to be concentrated in the plane $xz$ of the device, perpendicular to the axis of the conical electrodes. The electric field formed by the conical electrodes and the Faraday cup apparently pushes the particles into that plane. It can be also observed that only a fraction of the particles is captured by the Faraday cup (a simplified version of the Faraday cup was used in this model).

\begin{figure}[htb!]
  \centering
  \includegraphics[width=\columnwidth]{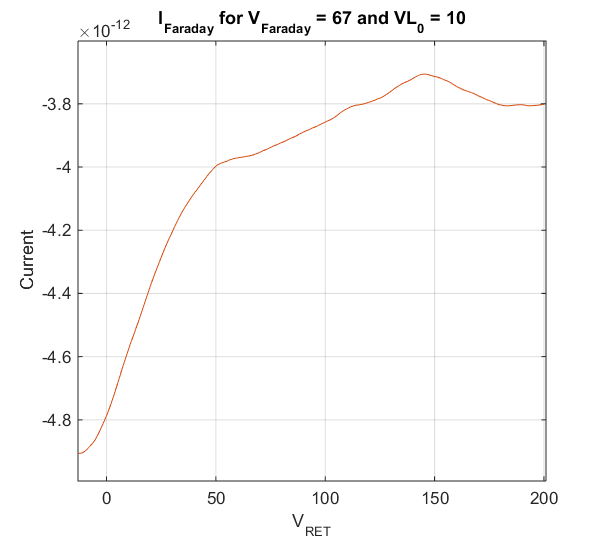}
  \caption{$I_{\textrm{Faraday}}$  vs. $V_{\textrm{C}}$ for $|V_{\textrm{Faraday}}| =$ 67 V and $V_{L_0} = 10$ V. }\label{Electrons}
\end{figure}

\begin{figure}[htb!]
  \centering
  \includegraphics[width=.9\columnwidth]{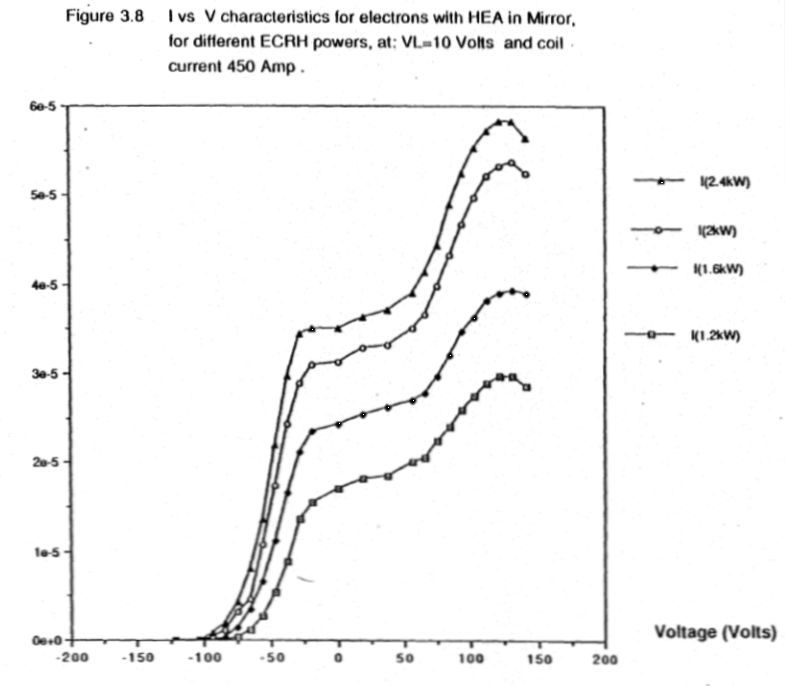}
  \caption{IV characteristic for only electrons, for $|V_{\textrm{Faraday}}| =$ 67 V and $V_{L_0} = 10$ V}\label{LealElectrons}
\end{figure}

\begin{figure}[htb!]
  \centering
  \includegraphics[width=.8\columnwidth]{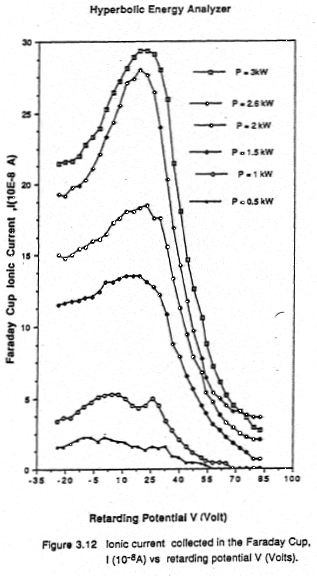}
  \caption{IV characteristic for only ions, for $|V_{\textrm{Faraday}}| =$ 67 V and $V_{L_0} = 10$ V}\label{LealIons}
\end{figure}

%Figure \ref{IVOnlyElectrons} shows the IV characteristic obtained when the particles in the system are only electrons. It may be noticed that the behavior of the current shown resembles the behavior of the currents obtained for electrons obtained in \cite{Le1989}.
Figure \ref{Electrons} shows the current collected at the Faraday Cup for a the distributions shown in Figure \ref{Input_and_Output_distributions}. It may be observed that the form of the IV characteristic for ions (left side of the plot) is very similar to the ones obtained in \cite{Le1989}. The right side of the plot shows the electron current, which is also similar to the behavior  reported in \cite{Le1989}, which is shown in Figure \ref{LealElectrons}. Notice that from -50 V to about 100 V the IV characteristic shows a behavior very similar to the one shown at Figure \ref{LealElectrons}. From -150 V to -50 V, the shape of the IV characteristic shows the same behavior as the ion current collected by Leal in \cite{Le1989},  shown in Figure \ref{LealIons}.

\begin{figure}[htb!]
  \centering
  \includegraphics[width=\columnwidth]{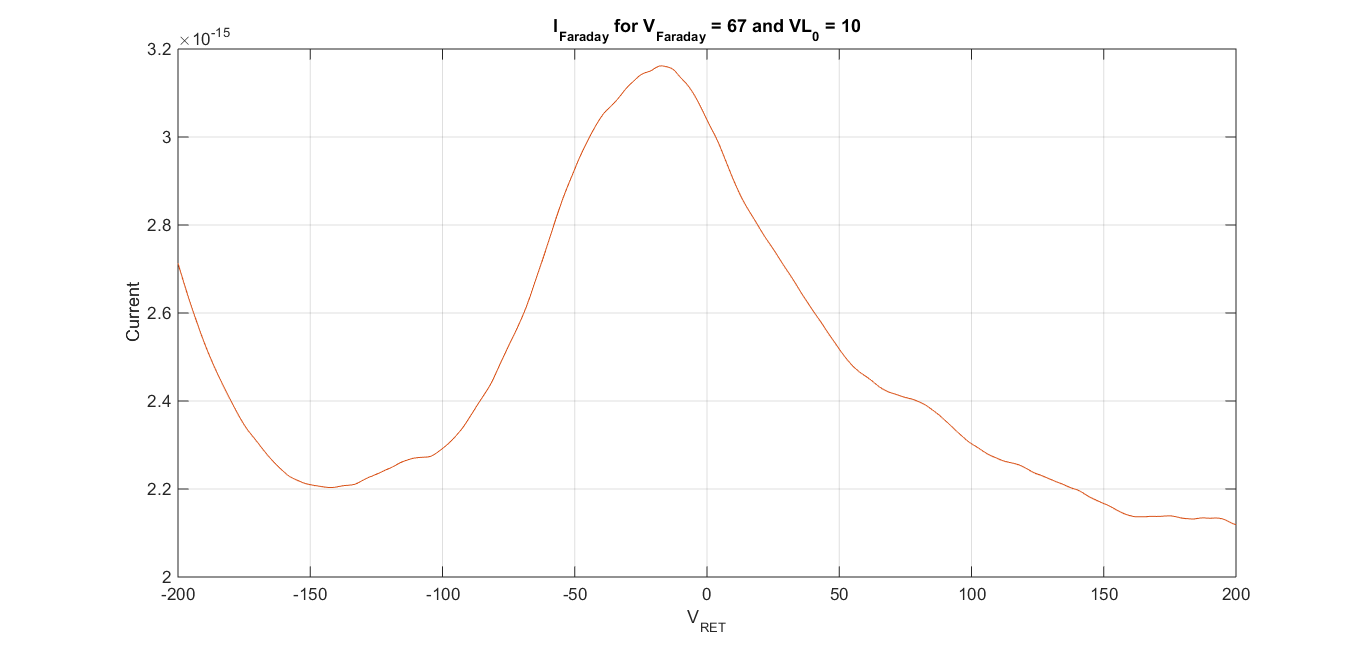}
  \caption{$I_{\textrm{Faraday}}$  vs. $V_{\textrm{C}}$ for $|V_{\textrm{Faraday}}| =$ 67 V and $V_{L_0} = 10$ V. }\label{Ions}
\end{figure}

\section{Conclusions}

A simulation of the Hyperbolic Field Energy Analyzer is performed with satisfactory results, achieving to determine and plot the dependency between the saddle point potential and the retarding potential applied to the conical elements that form the hyperbolic potential lens at the center of the device.

The current at the collector (Faraday Cup) is also plotted against the retarding potential applied to the conical lens, exhibiting results similar to the ones obtained experimentally \cite{LePreGa91,LEAL2005, Le1989}.

The energy distribution obtained in the Faraday cup remarkably resembles the energy distribution at the input port (L0) (Figure \ref{Input_and_Output_distributions}). It is clear that the measurements taken with the sensor will be done on a population which is representative of the input population. This simulation, however, only takes into account the particles already inside the sensor, after collimated by the lens L0. A more thorough simulation must be done to take into account the discrimination performed for the collimator lens. 

\section*{Future Work}
Determining the relationship between the voltage applied to the Faraday cup and ability of the sensor to focus particles into it is the next step in the simulation of the HFEA. Also, establishing the performance of the sensor at higher magnitudes of E will be investigated, new geometries for the electrodes, and simulation of the sensor inside a plasma simulation are tasks in the plans of the authors. These tasks will require higher computational resources and probably software other than only matlab. 
%\printbibliography
%\pagebreak

\bibliographystyle{ieeetran}
\bibliography{RESEARCHBIB} 

\begin{thebibliography}{1}
\providecommand{\url}[1]{#1}
\csname url@samestyle\endcsname
\providecommand{\newblock}{\relax}
\providecommand{\bibinfo}[2]{#2}
\providecommand{\BIBentrySTDinterwordspacing}{\spaceskip=0pt\relax}
\providecommand{\BIBentryALTinterwordstretchfactor}{4}
\providecommand{\BIBentryALTinterwordspacing}{\spaceskip=\fontdimen2\font plus
\BIBentryALTinterwordstretchfactor\fontdimen3\font minus
  \fontdimen4\font\relax}
\providecommand{\BIBforeignlanguage}[2]{{%
\expandafter\ifx\csname l@#1\endcsname\relax
\typeout{** WARNING: IEEEtran.bst: No hyphenation pattern has been}%
\typeout{** loaded for the language `#1'. Using the pattern for}%
\typeout{** the default language instead.}%
\else
\language=\csname l@#1\endcsname
\fi
#2}}
\providecommand{\BIBdecl}{\relax}
\BIBdecl

\bibitem{SSH1976}
W.~E. H. P.~B. Shyn, T. W.;~Sharp, ``Gridless retarding potential analyzer for
  use in very low energy charged particle detection,'' \emph{Review of
  Scientific Instruments}, vol.~47, no.~9, pp. 1005--1015, September 1976.

\bibitem{CZVRWHKBO2004}
Y.~Cui, Y.~Zou, A.~Valfells, M.~Reiser, M.~Walter, I.~Haber, R.~A. Kishek,
  S.~Bernal, and P.~G. O'Shea, ``Design and operation of a retarding field
  energy analyzer with variable focusing for space-charge-dominated electron
  beams", journal = {Review of Scientific Instruments}, year = {2004}, volume =
  {75}, number = {8}, pages = {2736-2745},.''

\bibitem{JGRA:JGRA8211}
\BIBentryALTinterwordspacing
D.~L. Reasoner, S.~D. Shawhan, and G.~Murphy, ``Plasma diagnostics package
  measurements of ionospheric ions and shuttle-induced perturbations,''
  \emph{Journal of Geophysical Research: Space Physics}, vol.~91, no. A12, pp.
  13\,463--13\,471, 1986. [Online]. Available:
  \url{http://dx.doi.org/10.1029/JA091iA12p13463}
\BIBentrySTDinterwordspacing

\bibitem{Le1989}
E.~Leal-Quir\'os, ``Novel probes and analyzers for {RF} heated plasmas and
  microwave heated plasmas for controlled fusion research: The hyperbolic
  energy analyzer, the magnetic moment analyzer, the double energy analyzer,
  and the variable energy analyzer,'' Ph.D. dissertation, University of
  Missouri-Columbia, 1989.

\bibitem{opac-b1133704}
I.~H. Hutchinson, \emph{Principles of plasma diagnostics}.\hskip 1em plus 0.5em
  minus 0.4em\relax Cambridge (U.K.), New York, Melbourne: Cambridge University
  Press, 2002, autre tirage : 2005.

\bibitem{opac-b1121477}
C.~K. Birdsall and A.~B. Langdon, \emph{Plasma physics via computer
  simulation}, ser. Series in plasma physics.\hskip 1em plus 0.5em minus
  0.4em\relax New York: Taylor \& Francis, 2005, originally published: New York
  ; London : McGraw-Hill, 1985.

\bibitem{LePreGa91}
M.~A.~P. E.~Leal-Quiros and E.~Garc\'ia-Otero, ``The hyperbolic energy
  analyzer: A novel diagnostic - ion probe,'' in \emph{Proceedings of the 10th
  International Workshop on ECR Ion Sources. . Conf.}, F.~Meyer and
  M.~Kirkpatrick, Eds.\hskip 1em plus 0.5em minus 0.4em\relax ORNL and US-DOE,
  January 1991, pp. 97--119.

\bibitem{LEAL2005}
\BIBentryALTinterwordspacing
E.~Leal-Quir\'os. (2005, June) Basic plasma diagnostic; probes \& analyzers.
  [Online]. Available: \url{http://www.pupr.edu/plasma}
\BIBentrySTDinterwordspacing

\end{thebibliography}

%\input{logo}

%\bibliographystyle{iopart-num}
%\bibliography{RESEARCHBIB}
\end{document}